\documentclass[aps,prd,a4paper,amsmath,amssymb,nofootinbib,showpacs,showkeys,
preprintnumbers,superscriptaddress,preprint]{revtex4-1}

\usepackage{latexsym,bm,amsmath,amssymb}
\usepackage[dvips]{color,graphicx}

\newcommand\keV{\mbox{keV}}

\newcommand\GeV{\mbox{GeV}}
\newcommand\TeV{\mbox{TeV}}
\newcommand\Pl{\mbox{\scriptsize Pl}}

\begin{document}

\title{Non-topological Domain Walls in a Model with Broken Supersymmetry}

\author{Leonardo Campanelli}\email{leonardo.campanelli@ba.infn.it}
\affiliation{Dipartimento di Fisica, Universit\`{a} di Bari,
I-70126 Bari, Italy}

\author{Marco Ruggieri}\email{ruggieri@yukawa.kyoto-u.ac.jp}
\affiliation{Yukawa Institute for Theoretical Physics,
 Kyoto University, Kyoto 606-8502, Japan\\
 {\bf Telephone number}: +81-75-753-7027}

\begin{abstract}

We study non-topological, charged planar walls (Q-walls) in the
context of a particle physics model with supersymmetry broken by
low-energy gauge mediation. Analytical properties are derived
within the flat-potential approximation for the flat-direction
raising potential, while a numerical study is performed using the
full two-loop supersymmetric potential. We analyze the energetics
of finite-size Q-walls and compare them to Q-balls,
non-topological solitons possessing spherical symmetry and arising
in the same supersymmetric model. This allow us to draw a phase
diagram in the charge-transverse length plane, which shows a
region where Q-wall solutions are energetically favored over
Q-balls. However, due to their finiteness, such finite-size
Q-walls are dynamically unstable and decay into Q-balls in a time
which is less than their typical scale-length.

\end{abstract}


~\footnote{The work of Marco Ruggieri is supported by JSPS with
contract number 09028.}

 \preprint{BA-TH-627-10}
\preprint{YITP-10-42} \maketitle


\section{Introduction}
Solitons are ``space-localized'' states which appear in certain
field theories (see~\cite{Lee:1991ax} for a review).
Depending on the nature of the boundary conditions, solitons can
be categorized as either topological or non-topological.

In the former case, the non-perturbative solution of the equation
of motions which corresponds to the soliton is characterized by a
topological charge; its stability is then guaranteed by the conservation
of the topological charge, which is usually zero for the vacuum and
non-zero for the soliton.

In the case of a non-topological
soliton~\cite{Lee:1974ma,Friedberg:1976me,Friedberg:1976eg,
Friedberg:1976az,Kobzarev:1974cp}, instead, the boundary
conditions are the same as those of the vacuum
state, while its stability, once
certain conditions are fulfilled, is
guaranteed by the conservation of a Noether charge associated to
an additional global symmetry of the action.

An interesting class of non-topological solitons is constituted by
spherical symmetric configurations known as Q-balls.
Firstly introduced by Coleman~\cite{Coleman:1985ki}, they were
subsequently studied by Kusenko~\cite{Kusenko:1997ad} who found
general conditions under which Q-balls are allowed as the ground state
of $U(1)$-charged scalar field theories.

Q-balls are solitons in more than one spatial dimension and
the formal environment in which they are understood is similar to that of vacuum
decay~\cite{Frampton:1976pb,Coleman:1977py,Callan:1977pt,Linde:1981zj}.
In order to avoid Derrick theorem~\cite{Derrick:1964ww}, Q-ball configurations
must be time-dependent. However, their properties can be made
stationary once a suitable time dependence of the solution is
imposed~\cite{Coleman:1985ki}:
\begin{equation}
\Phi(t,\bm r) = \frac{1}{\sqrt{2}} \, e^{i\omega t} \phi(r)~,
\label{eq:1}
\end{equation}
where $\phi(r)$ is the real part of the the complex scalar field $\Phi(t,\bm r)$
describing the Q-ball solution, and $\omega$ is a
real number which represents frequency of rotation in internal
$U(1)$ space. Spherical symmetry of the solution in the above
equation is kept manifest by writing the dependence on the
coordinates only by $r \equiv |\bm r|$.

There exist several studies about Q-balls, which analyze both their general properties~\cite{Q-balls-1,Copeland:2009as,Multamaki:1999an,Campanelli:2007um},
and implications for astrophysics and cosmology~\cite{Q-balls-2,Kusenko-Loveridge}.
In particular, the role of Q-balls in cosmology is enlightened in
Refs.~\cite{Kusenko,KusenkoTalk,Kusenko:1997si},
where it is shown that Q-balls arising in a supersymmetric particle physics model
where supersymmetry is broken by low-energy gauge mediation
are compelling candidate for baryonic dark-matter.
We will refer to such a kind of Q-balls as ``supersymmetric (SUSY) Q-balls''.

Moreover, Q-balls within the Signum-Gordon model have been analyzed in
Ref.~\cite{SIGNUM-GORDON}, while gravitational wave emission from the
fragmentation of a scalar condensate into Q-balls has been
investigated in Ref.~\cite{Kusenko:2008zm}. Finally, spinning
Q-balls in several contexts have been also investigated~\cite{SPINNING}.

In this article, we are mainly interested in studying solitonic solutions
which are less symmetric than spherical symmetric Q-balls.
They are characterized by planar symmetry in configuration space
and were first studied by MacKenzie and Paranjape~\cite{MacKenzie:2001av}:
They are known as Q-walls.

The interest of this study is twofold: Firstly, the existence of
solitonic configurations with higher symmetry, which are expected
to be those with lower energy, do not forbid the existence of
solitons with lower symmetry as excited states in dynamical
processes in which the solitons are involved, such as scattering,
fragmentation, evaporation, etc. Secondly, and to some extent also
surprisingly, since we will find that there exists a region in the
parameter space where finite-size, SUSY Q-walls solutions are
energetically more favored than SUSY Q-balls, we argue that not
always higher symmetry implies lower energy. Nevertheless, Q-walls
cannot be considered as the true ground state of the theory. As a
matter of fact, an analysis of finite-size effects reveals that
Q-walls, even when they are energetically favored over Q-balls,
decay in a finite amount of time, and the decay product are
Q-balls with the same value of charge.

The plan of the paper is as follows. In section II, we review
Q-ball solutions both within the context of a generic, complex scalar field
theory and in a specific model of supersymmetry.
In section III, we define and study Q-walls. We firstly derive
some general results, independent on the analytic form of the
potential. We then introduce Q-walls in the context of low-energy
gauge mediation SUSY breaking and derive the equation of state of
these solitons both in the flat-potential approximation and in
the full potential cases. In Section IV,
we address the stability of finite-size SUSY Q-walls,
and analyze their energetics with respect to that of SUSY Q-balls.
Finally, in Section V, we summarize our results and
draw our conclusions.


\section{Q-balls: An overview}

In this section, we briefly review both the Q-ball solution
introduced by Coleman in Ref.~\cite{Coleman:1985ki} and the
supersymmetric Q-ball configuration firstly discussed by Dvali,
Kusenko and Shaposhnikov~\cite{Dvali}.

\subsection{General Properties}

We consider a charged scalar field $\Phi$ whose lagrangian density
is given by
\begin{equation}
\label{eq:Lagr1} {\cal L} = \partial_\mu \Phi^* \partial^\mu\Phi -
U(|\Phi|)~.
\end{equation}
Here $U(|\Phi|)$ is a potential whose form will be specified later
for the case of SUSY Q-balls. For the moment, we simply require it
is invariant under a global $U(1)$ transformation. We normalize
the corresponding conserved Noether charge, $q$, as
\begin{equation}
\label{eq:charge} q = \frac{1}{i} \int \! d^3x \, (\Phi^*
\dot{\Phi} - \Phi \, \dot{\Phi}^*)~,
\end{equation}
where a dot indicates a derivative with respect to time. For a
given field configuration $\Phi$ the total energy is given by
\begin{equation}
\label{eq:energy} E = \int \! d^3x \left[|\dot{\Phi}|^2 + |\nabla
\Phi|^2 + U(|\Phi|) \right]~.
\end{equation}
In this paper, we are interested to solutions of the classical
field equations that correspond to time-independent total energy
$E$ and to a fixed value, namely $Q$, of the charge $q$ in
Eq.~\eqref{eq:charge}. Those requirements are satisfied by the
choice in Eq.~\eqref{eq:1} with $\omega$ considered as
a Lagrange multiplier associated to $Q$, and by the requirement that the
physical solitonic configuration renders the functional
\begin{equation}
\label{new} {\cal{E}}_\omega \equiv E + \omega (Q-q)
\end{equation}
stationary with respect to independent variations of $\Phi$,
$\Phi^*$ and $\omega$:
\begin{equation}
\label{1} \frac{\delta \mathcal{E}}{\delta\Phi} = 0, \;\;\;\;
\frac{\delta \mathcal{E}}{\delta\Phi^*} = 0, \;\;\;\; \frac{\delta
\mathcal{E}}{\delta\omega} = 0.
\end{equation}
The first two constraints lead to the equations of motion of the
field $\phi$, namely
\begin{eqnarray}
\label{eq:eqMotP} && \left( \nabla^2 + \omega^2 \right) \! \phi =
\frac{\delta U(\phi)}{\delta \phi}~,
\end{eqnarray}
while the latter one is equivalent to the requirement
that $\phi$ carries a total charge $q=Q$:
\begin{equation}
\label{q=Q} Q = \omega \! \int \! d^3x \, \phi^2~.
\end{equation}
In three spatial dimensions, it is usually assumed that the
profile function $\phi$ is isotropic, that is
$\phi = \phi(r)$. In this case, the field equation reads
\begin{equation}
\label{eq:eqMot3} \frac{d^2\phi}{dr^2} + \frac{2}{r}\frac{d
\phi}{d r} = \frac{\delta U(\phi)}{\delta \phi} - \omega^2 \phi~.
\end{equation}
Once we interpret the field $\phi$ as the particle position $x$
and $r$ as the time $t$, Eq.~\eqref{eq:eqMot3} is formally equal
to the equation of motion in one dimension of a particle moving in the
potential
\begin{equation}
\label{eq:UtildeTRE} \widetilde{U}(x) = -U(x) + \frac{1}{2} \,
\omega^2 x^2
\end{equation}
and subject to viscous damping. We will extensively use this
Newtonian analogy throughout this paper. In passing, we note that
the previous equations are formally similar to the equations for
the vacuum decays~\cite{Coleman:1977py,Callan:1977pt}.

It has been shown~\cite{Coleman:1985ki} that, given the boundary
conditions
\begin{equation}
\label{eq:Boundary} \left.\frac{d\phi}{dr}\right|_{r=0} = 0,
\;\;\;\; \phi|_{r \rightarrow \infty} = 0~,
\end{equation}
a solution to Eq.~\eqref{eq:eqMot3} exists, for $\omega$ in the
range $[\omega_0,m]$, where $\omega_0 \equiv \sqrt{2U(\phi_0)/\phi_0}$
and $\phi_0$ corresponds to the value of $\phi$ which minimizes
$\sqrt{2U(\phi_0)/\phi_0}$. Such a solution, which is an exact
solution of the classical equation of motion and represents a
non-topological soliton with charge $Q$, is said to be a Q-ball.
In the Newtonian interpretation, the Q-ball solution
corresponds to a particle starting at $t=0$ from a position $x_0$ with
velocity $v_0 = 0$ and reaching the origin in an infinite amount
of time.

In the field theory language, for each value of $\omega$
the Q-ball solution is equivalent to the bounce for tunneling in
three Euclidean dimensions in the potential
$\widetilde{U}$~\cite{Kusenko:1997ad,Kobzarev:1974cp,Frampton:1976pb,
Coleman:1977py,Callan:1977pt,Linde:1981zj}. Among the acceptable
values of $\omega$, the physical bounce solution is the one which
minimizes the functional ${\cal{E}}_\omega$ in Eq.~\eqref{new} for
fixed value $Q$ of the charge.

Once the Q-ball solution is found, one has to check its stability.
To this end, several kinds of stabilities must be kept into
account. First of all, as already pointed out by
Coleman~\cite{Coleman:1985ki}, the Q-ball could decay into a state
consisting of plane waves of the quanta of the field $\phi$ (that
constitute the perturbative spectrum of the theory). Since the
charge $Q$ is conserved, the Q-ball can decay only into a state
with the same value of the charge operator. The perturbative
spectrum consists of particles of mass $m$. Therefore, if the
energy of the Q-ball, given by Eq.~\eqref{eq:energy}, is lower
than $m Q$, then it can not decay into the quanta of the field.
This is the {\em absolute} stability requirement:
\begin{equation}
E < m Q~. \label{eq:ASta}
\end{equation}
Secondly, the Q-ball must be stable against quantum fluctuations.
A general powerful result there exists, which states
that stability against quantum fluctuations is achieved if the
equation of state of the Q-ball satisfies the condition:
\begin{equation}
\frac{\omega}{Q}\frac{dQ}{d\omega}\leq 0~. \label{eq:QSta}
\end{equation}
(For a proof of the above criterion see, e.g., the article by F.~Paccetti
Correia and M.~G.~Schmidt~\cite{Q-balls-1}, and Ref.~\cite{Copeland:2009as}.)
Equation~\eqref{eq:QSta} is commonly known as {\em classical}
stability criterion.

Finally, a Q-ball must be stable against fission into smaller
Q-balls. As a matter of fact, this process is not forbidden by
symmetry, as long as the total charge of the solitons in the final
state is equal to the charge of the Q-ball in the initial state.
It can be proved~\cite{Lee:1991ax} that the requirement of
{\em stability against fission} is equivalent to the following
condition:
\begin{equation}
\frac{d\omega}{dQ} < 0~. \label{eq:FSta}
\end{equation}
A comparison between Eq.~\eqref{eq:QSta} and~\eqref{eq:FSta} shows
that quantum stability requirement is a less stringent
constraint with respect to stability against fission. However, in
the whole regime in which our results are physically relevant, the
relation $\omega=\omega(Q)$ is always monotonous. As a
consequence, in our context the two stability requirements are
equivalent to each other.

\subsection{Supersymmetric Q-balls}

In this subsection, we summarize the main properties of Q-balls
arising in a model with broken supersymmetry. In the present
model~\cite{SUSY}, known as ``gauge-mediation SUSY breaking'', a
coupling between vector-like messenger fields and ordinary gauge
multiplets breaks explicitly supersymmetry, the coupling constant
among the two kinds of field being of order $g \sim 10^{-2}$.
Because of this coupling, the potential for a generic flat
direction $\phi$ does not longer vanish; instead, it is lifted up
by an amount which has been computed at the lowest non-vanishing
(two-loop) order in Ref.~\cite{de Gouvea}:
\begin{equation}
\label{potential} U(\chi) = \Lambda \!\! \int_0^1 \!\! dx \,
\frac{\chi^{-2} - x(1-x) +
x(1-x)\ln[x(1-x)\chi^2]}{[\chi^{-2}-x(1-x)]^2}~.
\end{equation}
Here, $\chi \equiv \phi/M$ and $M \equiv M_S/(2g)$, with $M_S$ the
messenger mass scale. The value of the mass parameter
$\Lambda^{1/4}$ is constrained as (see, e.g., Ref.~\cite{Kasuya511}):
$10^3 \GeV \lesssim \Lambda^{1/4} \lesssim (g^{1/2}/4\pi)
\sqrt{m_{3/2} M_{\Pl}}$,
where $M_{\Pl} \sim 2.4 \times 10^{18} \GeV$ is the reduced Planck
mass and the gravitino mass, $m_{3/2}$, is in the range
$100 \, \keV \lesssim m_{3/2} \lesssim 1\GeV$~\cite{de Gouvea,Kasuya511}.

The asymptotic expressions of $U(\chi)$, for small and large
$\chi$ are~\cite{de Gouvea}:
\begin{equation}
\label{potentialapprox} \frac{U(\chi)}{\Lambda} \simeq
    \left\{ \begin{array}{ll}
        \chi^2, &  \;\; \mbox{if} \;\; \chi \ll 1, \\
        (\ln \chi^2)^2 - 2 \ln \chi^2 + \frac{\pi^2}{3} \, , &  \;\; \mbox{if} \;\; \chi \gg 1.
    \end{array}
    \right.
\end{equation}
In the context of gauge-mediation SUSY breaking, a widely used
approximation in studying Q-balls consists in replacing the full
potential $U(\chi)$ with its asymptotic expansions~\eqref{potentialapprox},
in which a plateau plays the role of the logarithmic rise for large
values of $\chi$:
\begin{equation}
\label{potentialapprox2} U(\phi) =
    \left\{ \begin{array}{ll}
            \frac{1}{2} \, m^2 \phi^2, &  \phi \leq M, \\
            \Lambda, &  \phi \geq M ,
    \end{array}
    \right.
\end{equation}
where $m \equiv \sqrt{2 \Lambda}/M$ is the soft breaking mass
which is of order $1\TeV$~\cite{Kasuya511}. We will refer to such
an approximation as the {\em flat-potential approximation}.
It has been shown that the approximated potential $U(\phi)$ allows
Q-balls solutions as the non perturbative ground state of the
model~\cite{Dvali,Kusenko-Loveridge}. Such states are known as
SUSY Q-balls.

In the limit of large charges, one can compute analytically the
most important characteristics of SUSY Q-balls. In particular,
the Q-ball energy is given by~\cite{Dvali,Kusenko-Loveridge}:
\begin{equation}
\label{Eball1} \frac{E}{m Q_{\rm cr}} \simeq \frac{4 \sqrt{2}
\pi}{3} \left( \frac{Q}{Q_{\rm cr}} \right)^{\!3/4} \!,
\end{equation}
valid for $Q \gg Q_{\rm cr}$, where
\begin{equation}
\label{criticalQ} Q_{\rm cr} \equiv \Lambda/m^4
\end{equation}
is the so-called critical charge.

Taking into account the full potential~\eqref{potential}, only
numerical calculations are feasible. In this case, one finds
absolutely stable Q-ball solutions, $E/mQ < 1$, only if the charge
of the soliton is larger than $Q_{\rm min}$,
where~\cite{Campanelli:2007um}
\begin{equation}
\label{Qmin} Q_{\rm min} \simeq 504 Q_{\rm cr}.
\end{equation}
Moreover, the exact relationship between the energy of a SUSY
Q-ball and its charge, can be written as
\begin{equation}
\label{Eball2} \frac{E}{m Q_{\rm cr}} = \xi_E(Q)
\left(\frac{Q}{Q_{\rm cr}}\right)^{\! 3/4} \!,
\end{equation}
where $\xi_E$ is a slowly increasing function of $Q$, computed
in~\cite{Campanelli:2007um}. This function can be fitted by a simple
analytical form, namely
\begin{equation}
\label{xi} \xi_E(Q) = a + b \log_{10}^p (Q/Q_{\rm cr}),
\end{equation}
with $a \simeq -17.438$, $b \simeq 15.559$, and $p \simeq 0.352$.
The maximum percentage error of the function $\xi_E$ with respect
to its numerical value is less than $2.8\%$ for $Q$ in the range
$Q \in [Q_{\rm min}, 7.2 \times 10^{37} Q_{\rm cr}]$.


\section{Q-walls}

\subsection{General Properties}

Next we turn to the main object of our study, namely
Q-walls. We define a Q-wall as a solitonic solution of the equation of motion
possessing planar symmetry, namely such that $\phi({\bf r}) = \phi(z)$,
where $z$ is the coordinate perpendicular to the wall. The field
equation~\eqref{eq:eqMotP} for the wall profile $\phi$ reads
\begin{equation}
\label{eq:eqMot} \frac{d^2 \phi}{dz^2} = \frac{\delta
U(\phi)}{\delta \phi} - \omega^2 \phi~,
\end{equation}
and it must be solved with the following boundary conditions
\begin{equation}
\label{eq:BoundaryWall} \left.\frac{d\phi}{dz}\right|_{z=0} =
0~,~~~~~\phi|_{z \rightarrow \pm \infty} = 0~,
\end{equation}
analogous to the Coleman's boundary conditions for the Q-balls. If
we interpret $\phi$ as a coordinate $x$ and $z$ as the time $t$,
then Eq.~\eqref{eq:eqMot} is formally equal to the equation of
motion of a particle moving in the effective potential
\eqref{eq:UtildeTRE}. The force acting on the particle, $F =
-\partial\widetilde{U}/\partial x$, is conservative. In the case
of spherical symmetric profile studied in the previous section, a
viscous term was present. The difference among the two cases is a
trivial consequence of the analytical form of the Laplacian
operator in one and three spatial dimensions, respectively. As a
consequence of the conservative nature of the effective force
acting on our fictitious particle, for any given initial condition
$x(t=0)\equiv x_0$, $(dx/dt)_{t=0} \equiv v_0$, the solution of
the equation of motion $x(t)$ can be found by means of the
quadrature of the first integral of energy:
\begin{equation}
\label{eq:solt} t = \pm \sqrt\frac{m}{2} \int_{x_0}^x
\frac{dy}{\sqrt{E_{\rm tot} - \widetilde{U}(y)}}~,
\end{equation}
where $m$ is the mass of the particle and
$E_{\rm tot} = m v_0^2/2 + \widetilde{U}(x_0)$.
Translating Eq.~\eqref{eq:solt} to our field
theoretical problem, we can write the solution $\phi(z)$ of
Eq.~\eqref{eq:eqMot} in implicit form as
\begin{equation}
\label{eq:solImpl} z = \pm \sqrt\frac{1}{2}\int_{\phi_0}^\phi
\frac{dy}{\sqrt{\widetilde{U}(\phi_0)-\widetilde{U}(y)}}~,
\end{equation}
where $\phi_0 \equiv \phi(z=0)$. If we normalize the potential in such a way that $U(0)=0$,
then the quantity $\phi_0$ satisfies the condition
\begin{equation}
\label{condition-phi0} \widetilde{U}(\phi_0) = 0.
\end{equation}
Our initial conditions correspond, in the problem of the motion of
a particle in the potential $\widetilde{U}$, to solutions that
describe the motion of the particle starting at $t=0$ from some
position $x_0$ with zero velocity $v_0$ and arriving to the point
$x=0$ at $t = +\infty$ with velocity $v_\infty = 0$.

For a generic potential $U$, we can argue
the form of the Q-wall profile by looking at the
approximate solutions of the full equation of motion, in the limit
of small or large $z$. For small $z$ one can
write $\phi(z) \simeq \phi_0 + c z^2$ [the condition
$\phi^\prime(z=0) = 0$ has to be satisfied, see
Eq.~\eqref{eq:BoundaryWall}]. Substituting into Eq.~\eqref{eq:eqMot} we find
$c = -(1/2) (\partial{\widetilde{U}}/\partial \phi)|_{\phi_0}$,
Moreover, for large $z$ one has $\phi \rightarrow 0$ and can
keep only the quadratic terms in the effective potential
$\widetilde{U}(\phi) \simeq (\omega^2 - m^2)\phi^2/2$. In this case,
the equation of motion~\eqref{eq:eqMot} linearizes and the
solution is easily found to be $\phi(z) \simeq \phi_0
e^{-z\sqrt{m^2 - \omega^2}}$. Summarizing, the asymptotical
behavior of the wall profile is given by
\begin{eqnarray}
\frac{\phi(z)}{\phi_0} =
    \left\{ \begin{array}{ll}
            1 - \frac12 \frac{\partial \widetilde{U}}{\partial \phi}|_{\!\phi_0} \: z^2, &  z \rightarrow 0,
            \\ \\
            e^{-z\sqrt{m^2 - \omega^2}}, &  z \rightarrow +\infty.
    \end{array}
    \right.
\label{eq:profileZZ}
\end{eqnarray}
Equation~\eqref{eq:profileZZ} defines a decay length,
\begin{equation}
\Delta = \frac{1}{\sqrt{m^2 - \omega^2}} \, , \label{eq:DELTA}
\end{equation}
that we identify with the {\em thickness} of the wall.

As in the case of Q-balls, the existence of Q-walls is subject to
the absolute stability condition Eq.~\eqref{eq:ASta}. Due to
planar symmetry of the solitonic solution, it is convenient to introduce the
surface densities of charge and energy, $\sigma$ and $\rho$,
respectively as
\begin{equation}
\label{sigmarhoDef} Q \equiv A \sigma, \;\;\;\; E \equiv A \rho,
\end{equation}
where $A \equiv \int d^2x$ is the surface area of the wall, and
\begin{eqnarray}
&& \rho = \int \! dz \! \left[\frac{1}{2} \, \omega^2 \phi^2 +
\frac{1}{2} \, (\partial_z \phi)^2 + U(\phi) \right] \! ,
\label{eq:RHO}\\
&& \sigma = \omega \! \int \! dz \phi^2.
\end{eqnarray}
Using definition~\eqref{sigmarhoDef}, the absolute stability
condition reads
\begin{equation}
\rho < m \sigma.
\end{equation}
In the next section, we will focus on stable Q-walls in a scalar
field theory defined by the supersymmetric
potential~\eqref{potential}. Before discussing the specific case,
it is interesting to note here that, for any form of the potential in
Eq.~\eqref{eq:RHO}, the surface energy and charge of a Q-wall are
connected by the following parametric equation of state:
\begin{eqnarray}
\label{parametric1} \rho(\omega) \!\!&=&\!\!  S(\omega) +\omega\sigma(\omega), \\
\label{parametric2} \sigma(\omega) \!\!&=&\!\! -\frac{dS}{d\omega}
\, ,
\end{eqnarray}
where
\begin{equation}
\label{S} S(\omega) \equiv  2\sqrt{2}\int_0^{\phi_0} \!\! d\phi \,
\sqrt{-\widetilde{U}(\phi)} \, ,
\end{equation}
and $\phi_0$ is the solution of Eq.~\eqref{condition-phi0}.
Equations~\eqref{parametric1} and~\eqref{parametric2} are nothing
but the statement that the total energy density of the Q-wall is
the Legendre transform of the action integral $S(\omega)$. For a
given form of the potential $U$, one can eliminates the $\omega$
parameter between Eqs.~\eqref{parametric1} and
\eqref{parametric2}, and can then express the surface energy
$\rho$ as a function of the surface charge $\sigma$. The resulting
equation of state can be finally used to check the stability
condition.

\subsection{Supersymmetric Q-walls: Analytical results}

In this subsection, we derive analytical properties of
supersymmetric Q-walls. This is possible when considering the
simplified case of flat potential~\eqref{potentialapprox2}. In the
context of Q-balls, it was shown~\cite{Campanelli:2007um}
that the using of the full potential simply induces logarithmic
corrections to the power-law relationships among the various
physical quantities of Q-ball. We will show
later that this statement holds for Q-wall solutions as well.

Within the flat-potential approximation, we easily obtain the
expressions for the surface energy and the charge:
\begin{eqnarray}
\label{rhox}   && \rho = m \sigma h(\omega/m), \\
\label{sigmax} && \sigma = \sigma_{\rm cr} \, g(\omega/m),
\end{eqnarray}
where
\begin{eqnarray}
\label{h} && h(x) \equiv x \left( 1 + \frac{1}{1 + \frac{x}{\sqrt{1-x^2} \arccos x}} \right) \!, \\
\label{g} && g(x) \equiv \frac{1}{x^2} \left( \frac{x}{\sqrt{1-x^2}} + \arccos x \right) \!,
\end{eqnarray}
The introduction of the critical charge,
\begin{equation}
\sigma_{\rm cr} \equiv M^2,
\end{equation}
will be useful in the following.

\begin{figure}[t]
\begin{center}
\includegraphics[clip,width=0.47\textwidth]{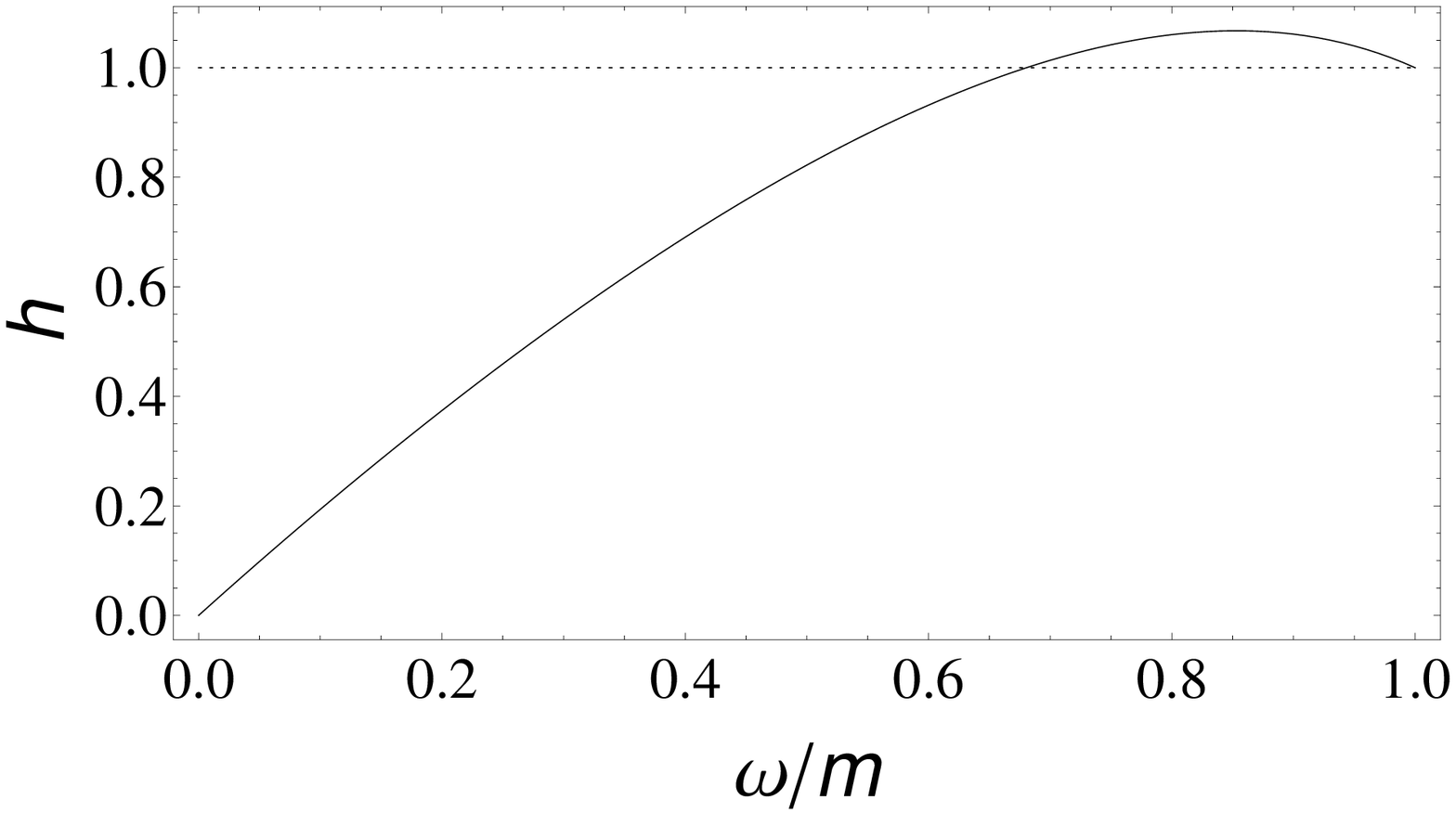}
\includegraphics[clip,width=0.47\textwidth]{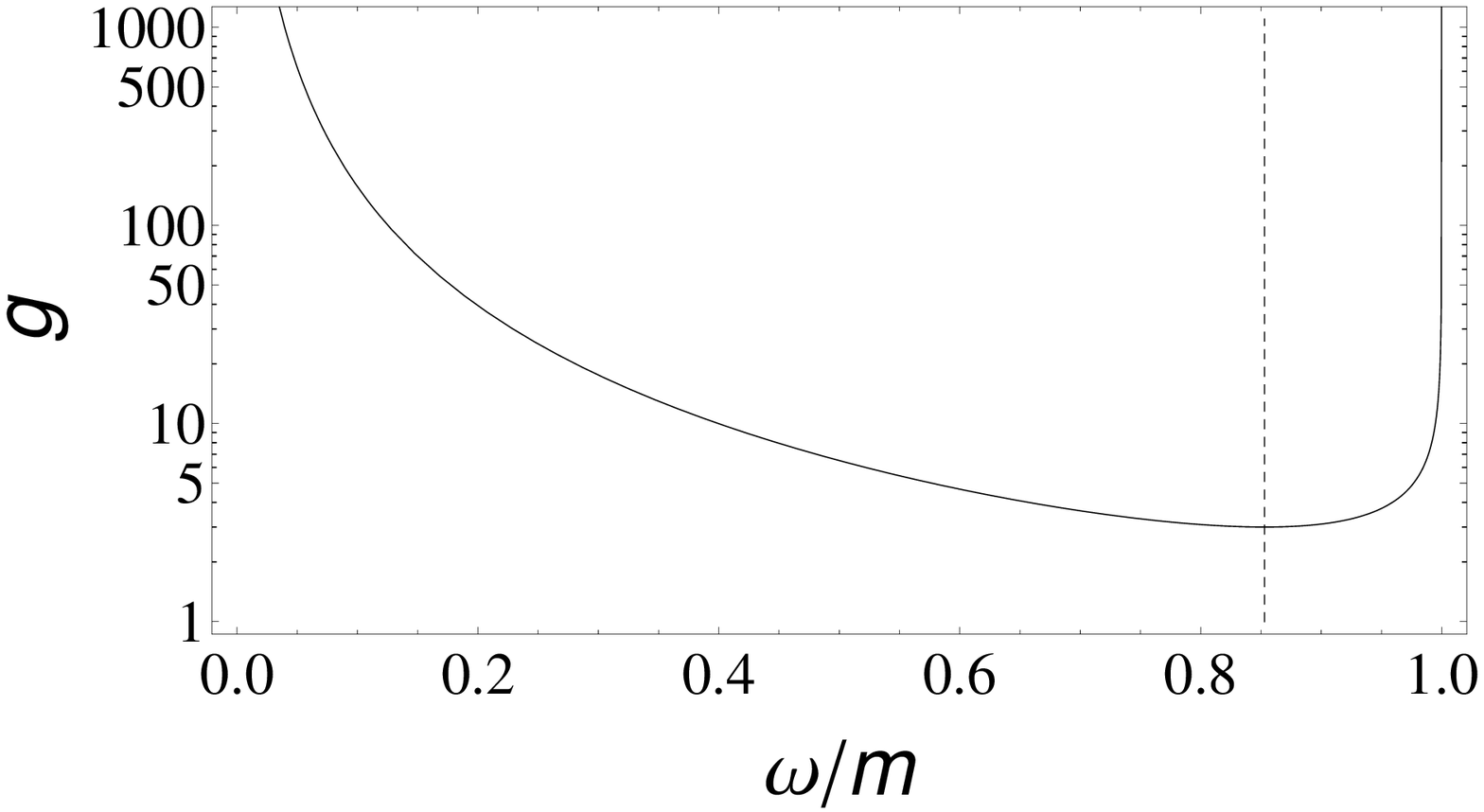}
\caption{\label{Fig:1}{\em Upper panel}. The function $h = \rho/m
\sigma $ as a function of $\omega$. When $h > 1$, SUSY Q-walls are
absolutely unstable and decay into particles of mass $m$.
{\em Lower panel}. The function $g = \sigma/\sigma_{\rm cr}$
against $\omega$. The vertical
dashed line corresponds to $\omega \simeq 0.85 m$. For charges larger
than this critical value of $\omega$, Q-walls are unstable
against fission and against quantum fluctuations.}
\end{center}
\end{figure}

We plot the function $h(\omega/m)$ in the upper panel of
Fig.~\ref{Fig:1}. Since $h = \rho/m \sigma$, the absolute
stability condition requires $h < 1$. Numerically, we find that
the above condition is fulfilled for $\omega \lesssim 0.68 \, m$.
Using Eq.~\eqref{sigmax} we then find that there exists a minimal
charge $\sigma_{\rm min} \simeq 3.00 \, \sigma_{\rm cr}$ such
that, for $\sigma < \sigma_{\rm min}$, SUSY Q-wall are classically
unstable and decay into particles of mass $m$. For completeness, in
the lower panel of Fig.~\ref{Fig:1} we plot the function
$g(\omega/m)$. It is interesting to note that there exists a
critical value of $\omega$, namely $\omega \simeq 0.85 m$, above
which Q-walls are unstable against quantum fluctuations and
fission (see Eq.~\eqref{eq:QSta}). However, this
is not relevant for our discussion, since in the region of
quantum and fission instability,  Q-walls are already unstable
against decay into free quanta.


\begin{figure}[t]
\begin{center}
\includegraphics[clip,width=0.47\textwidth]{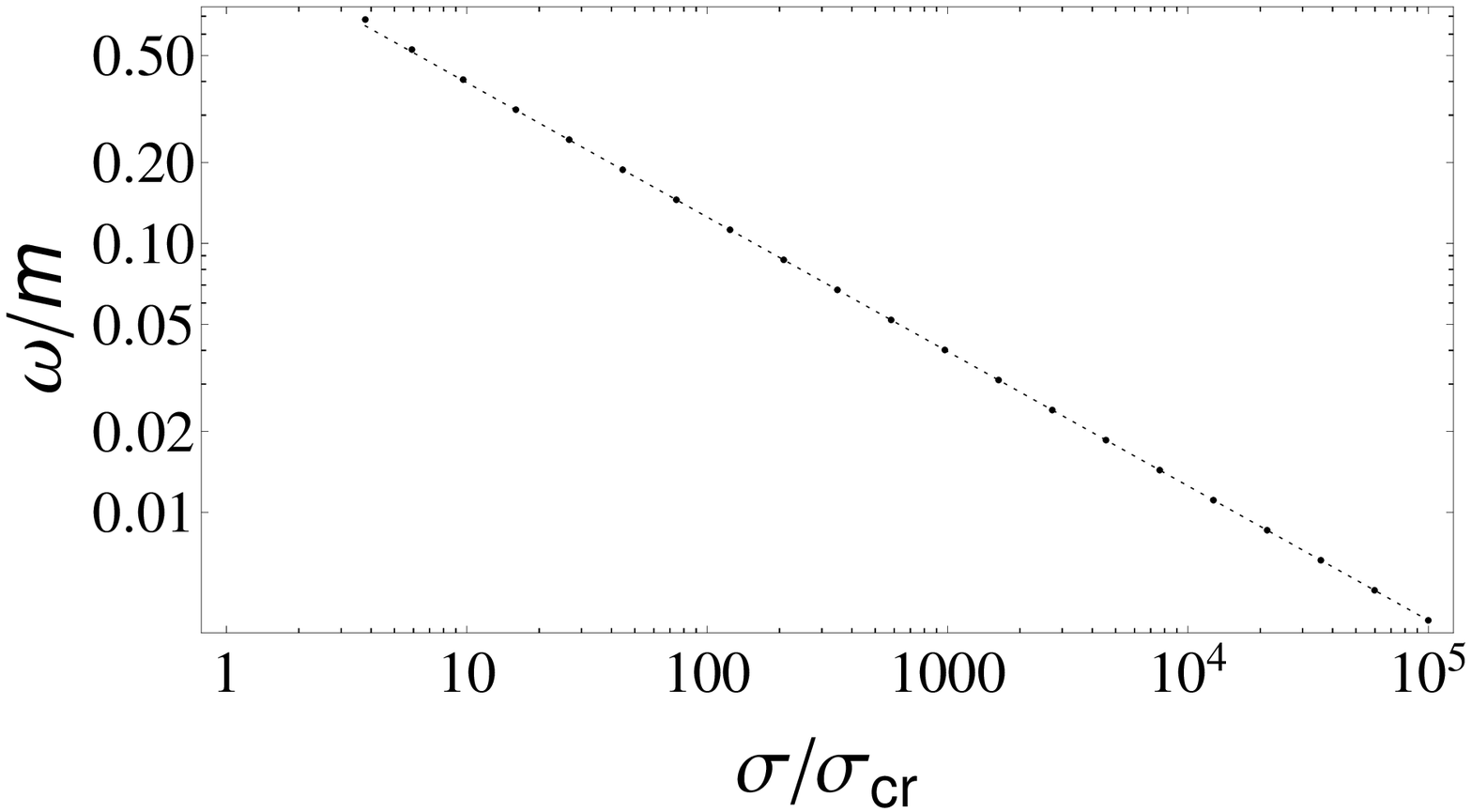}
\includegraphics[clip,width=0.47\textwidth]{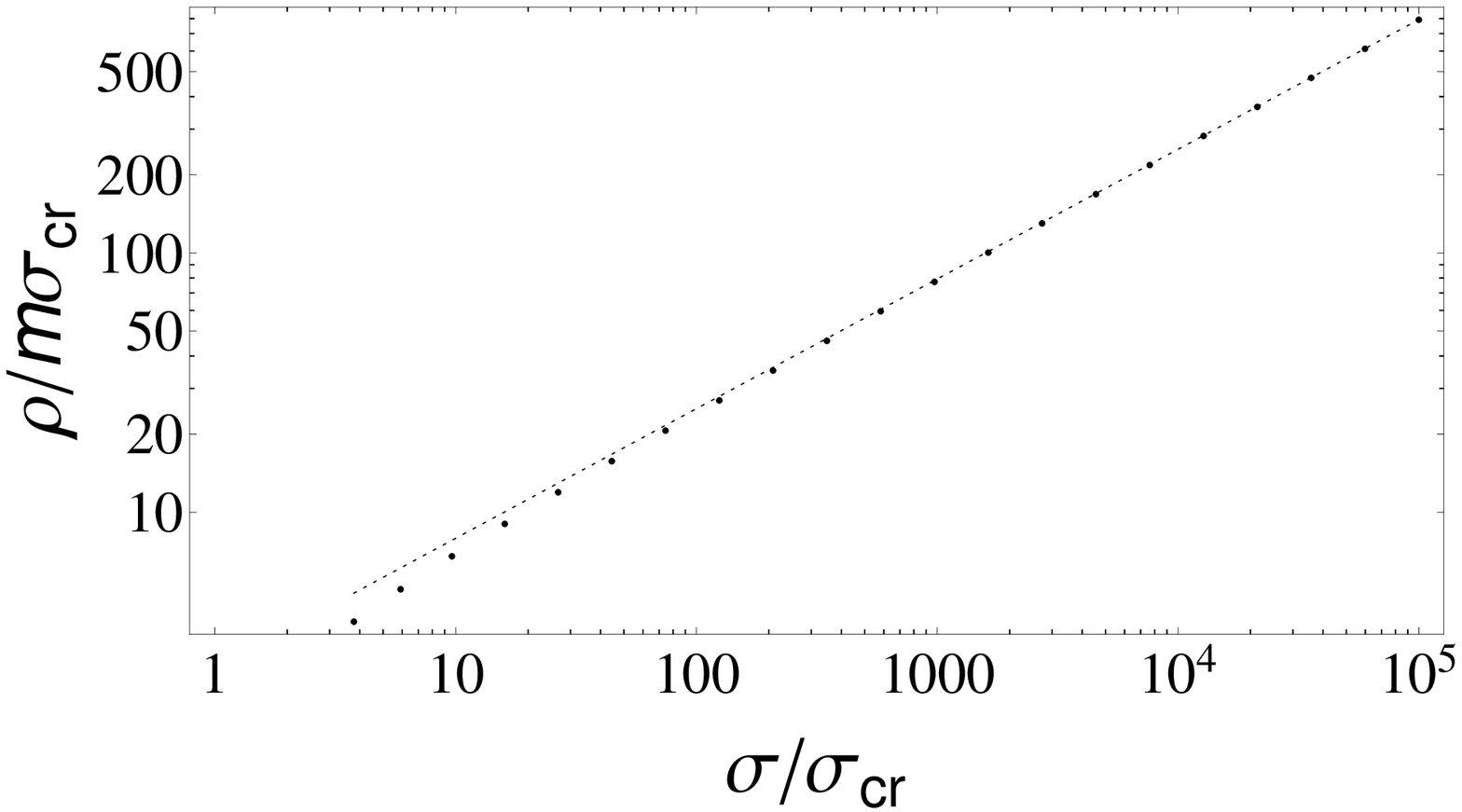}
\caption{The parameter $\omega$ (upper panel) and the surface
energy density $\rho$ (lower panel) as a function of the surface
charge $\sigma$. Dotted lines represent the analytical expressions
of $\omega$ and $\rho$ for large values of the charge.}
\end{center}
\end{figure}


\begin{figure}[t]
\begin{center}
\includegraphics[clip,width=0.47\textwidth]{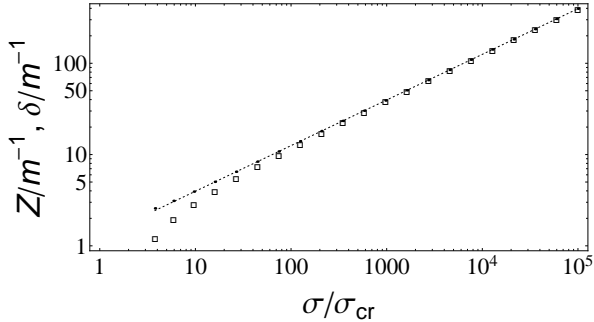}
\caption{The total width, $\delta$ (points), and the spread, $Z$
(empty squares), of a SUSY Q-wall as a function of the charge,
together with their analytical expressions for large charges
(dotted line). The difference between $\delta$ and $Z$ is the thickness of the
wall $\Delta$, which goes to zero for large charge.}
\end{center}
\end{figure}


\begin{figure}[t]
\begin{center}
\includegraphics[clip,width=0.47\textwidth]{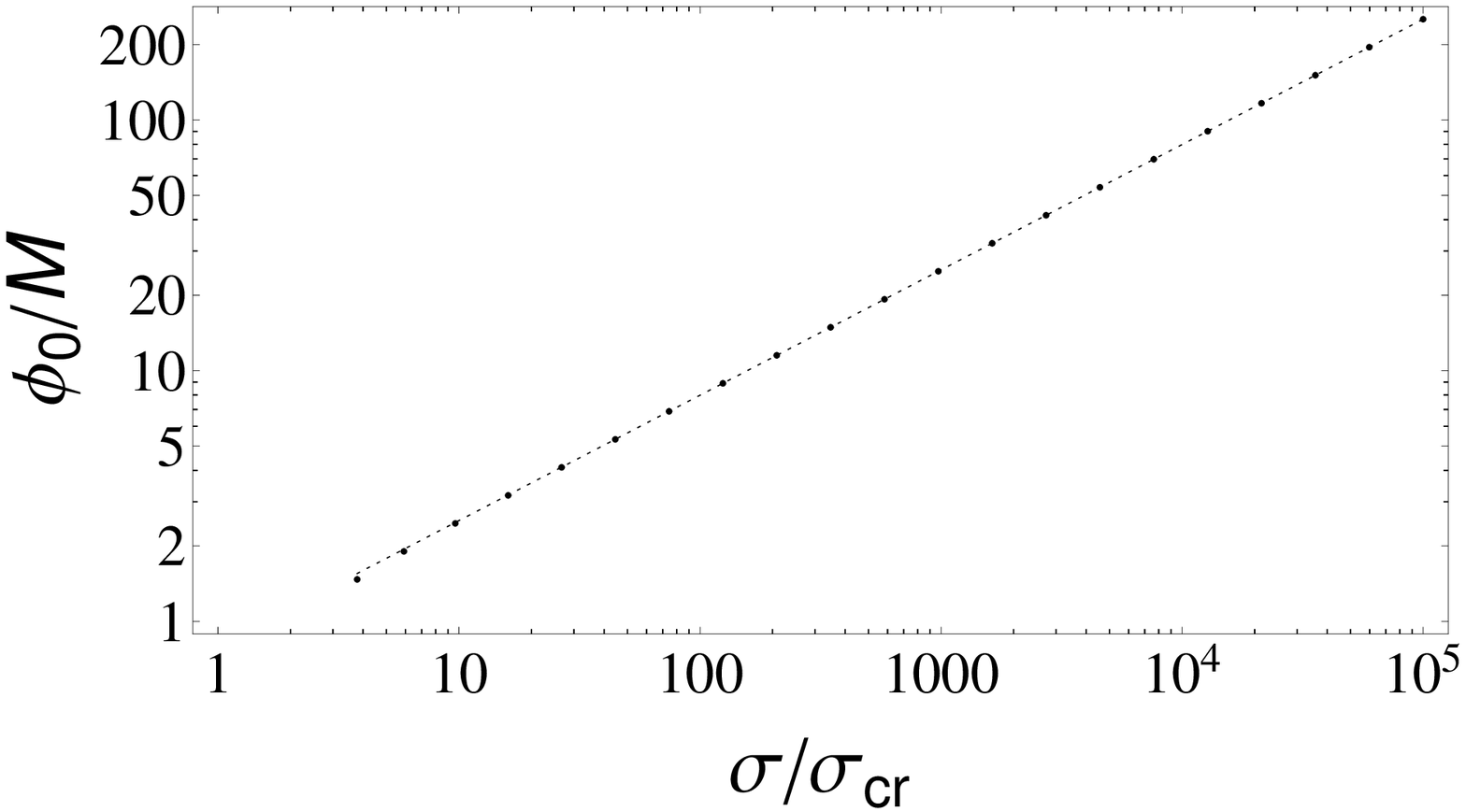}
\includegraphics[clip,width=0.47\textwidth]{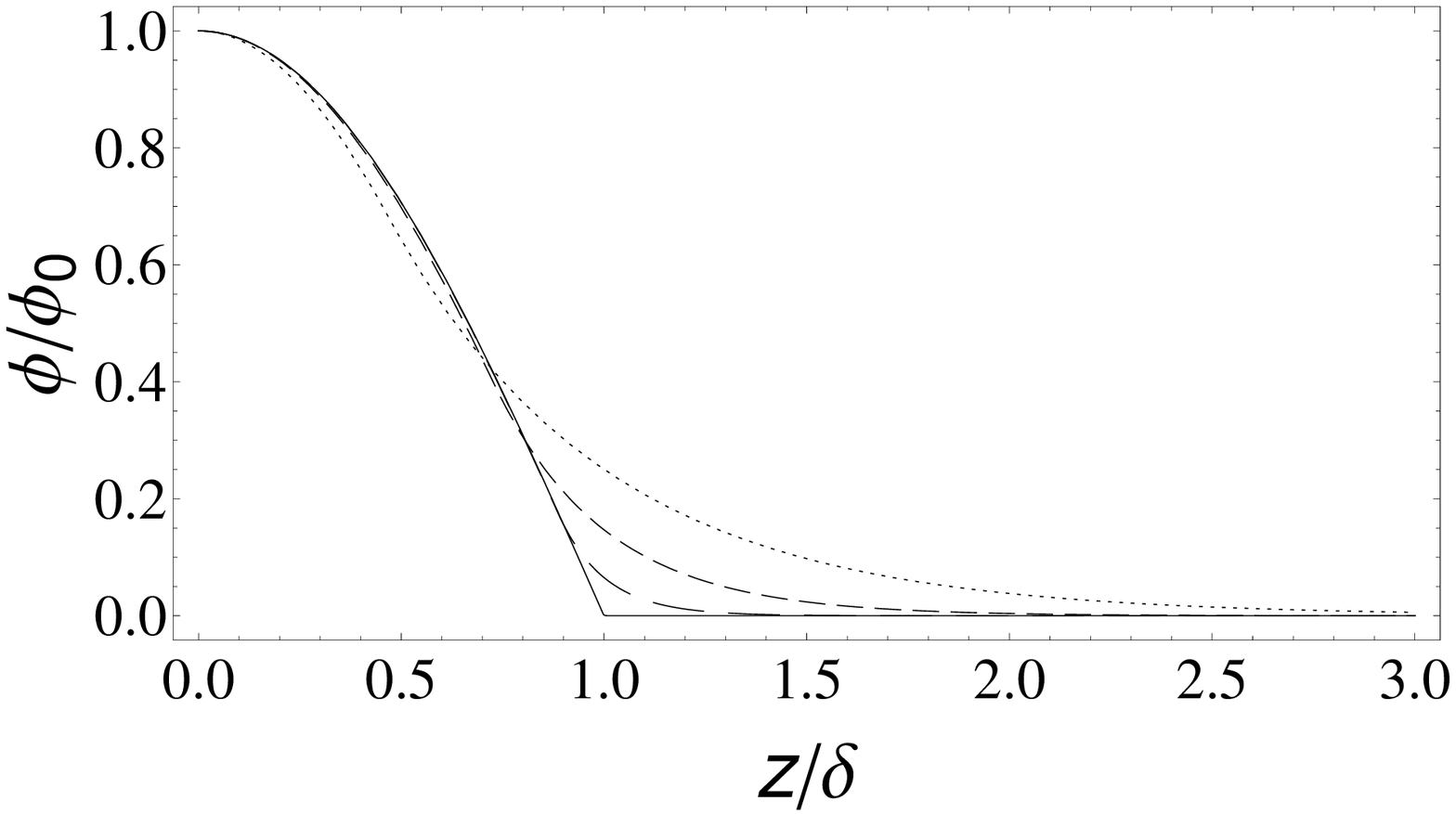}
\caption{{\it Upper panel.} $\phi_0$ as a function of the charge
(points) with its analytical expression for large charges (dotted
line). {\it Lower panel.} Susy Q-wall profile for different values
of the charge: $\sigma = 10^5 \sigma_{\rm cr}$ (continuous line),
$\sigma = 50 \sigma_{\rm cr}$ (long dashed line) $\sigma = 10
\sigma_{\rm cr}$ (dashed line), $\sigma = \sigma_{\rm min} \simeq
3.00 \sigma_{\rm cr}$ (dotted line).}
\end{center}
\end{figure}


In Fig.~2, we plot $\omega$ and $\rho$ versus the
charge  $\sigma$. The values of $\omega$ (which we indicate as points in the
plots) are found numerically by solving Eq.~\eqref{sigmax} as a
function of $\omega$. The corresponding solutions are then
inserted in Eq.~\eqref{rhox} to find the equation of state of a
Q-wall, that is $\rho$ as a function of $\sigma$. Dotted lines
represent the analytical expressions of $\omega$ and $\rho$ for
large values of the charge ($\sigma \gg \sigma_{\rm cr}$):
\begin{eqnarray}
\label{omegaA} && \frac{\omega}{m} \simeq
\left(\frac{\pi}{2}\right)^{\!1/2} \left(
\frac{\sigma}{\sigma_{\rm cr}} \right)^{\!-1/2} \!,
\\
\label{rhoA} && \frac{\rho}{m\sigma_{\rm cr}} \simeq (2\pi)^{1/2}
\left( \frac{\sigma}{\sigma_{\rm cr}} \right)^{\!1/2} \!,
\end{eqnarray}
which are in agreement with the results of Ref.~\cite{MacKenzie:2001av}.

Next we compute the Q-wall profile. From Eq.~\eqref{eq:solImpl} we
find:
\begin{equation}
\label{} \frac{\phi(z)}{\phi_0} =
    \left\{ \begin{array}{ll}
            \cos (\omega z), &  z \leq Z,
            \\ \\
            \cos (\omega Z) \, e^{-(z-Z)/\Delta}, &  z \geq Z,
    \end{array}
    \right.
\end{equation}
where the wall {\em spread}, $Z$, is defined by the condition $\phi(Z) = M$,
with $\Delta$ defined in Eq.~\eqref{eq:DELTA}. Solving
Eq.~\eqref{condition-phi0} we find $\phi_0 = m M/\omega$ and in
turns $Z = (1/\omega) \arccos (\omega/m)$.

The analytic form of the Q-wall profile in the above equation
suggests the introduction of the wall {\em width}, that we
denote by $\delta$, as
\begin{equation}
\label{deltaDef} \delta \equiv Z + \Delta~.
\end{equation}
In Fig.~3, we show $\delta$ (points in the plot) and $Z$ (empty
squares in the plot) as a function of the charge, together with
their analytical expressions for large charges (dotted line):
\begin{equation}
\label{deltaA} \frac{\delta}{m^{-1}} \simeq \frac{Z}{m^{-1}} \simeq
\left(\frac{\pi}{2}\right)^{\!1/2} \left( \frac{\sigma}{\sigma_{\rm cr}} \right)^{\!1/2} \!.
\end{equation}
It is worth noting that the limit of large charges is equivalent
to neglect the thickness of the wall with respect to its spread.

Finally, we plot in Fig.~4 the quantity $\phi_0$ as a function of
the charge (points in the plot), and compare it with its
analytical expression for large charges (dotted line):
\begin{equation}
\label{phi0A} \frac{\phi_0}{M} \simeq
   \left(\frac{2}{\pi}\right)^{\!1/2} \left( \frac{\sigma}{\sigma_{\rm cr}} \right)^{\!1/2} \!.
\end{equation}
In the same figure, we plot the Q-wall profile for different values
of the charge.

\subsection{Supersymmetric Q-walls: Full Potential}

We now present our results for the full
potential~\eqref{potential}. In analogy with the approximate case
discussed in the previous section, we define the quantity
\begin{equation}
h(\omega) \equiv \rho/m \sigma~, \label{eq:rhoxF}
\end{equation}
(compare with Eq.~\eqref{rhox}). We find numerically that absolute
stability condition, $h < 1$, is fulfilled for values of $\omega$ less
than a maximal value
\begin{equation}
\label{omegamax} \omega_{\rm max} \simeq 0.93 m~
\end{equation}
(see Fig.~5). This in turn means that there exists a minimum charge,
above which SUSY Q-walls are unstable and than decay into the quanta
of the field. Numerically we find
\begin{equation}
\label{sigmamin} \sigma_{\rm min} \simeq 1.23 \sigma_{\rm cr}.
\end{equation}
(It is now clear why we introduced the critical charge: It turns
to be, approximatively, the value of the charge deining the
limit between stable and unstable Q-walls.)

\begin{figure}[t]
\begin{center}
\includegraphics[clip,width=0.47\textwidth]{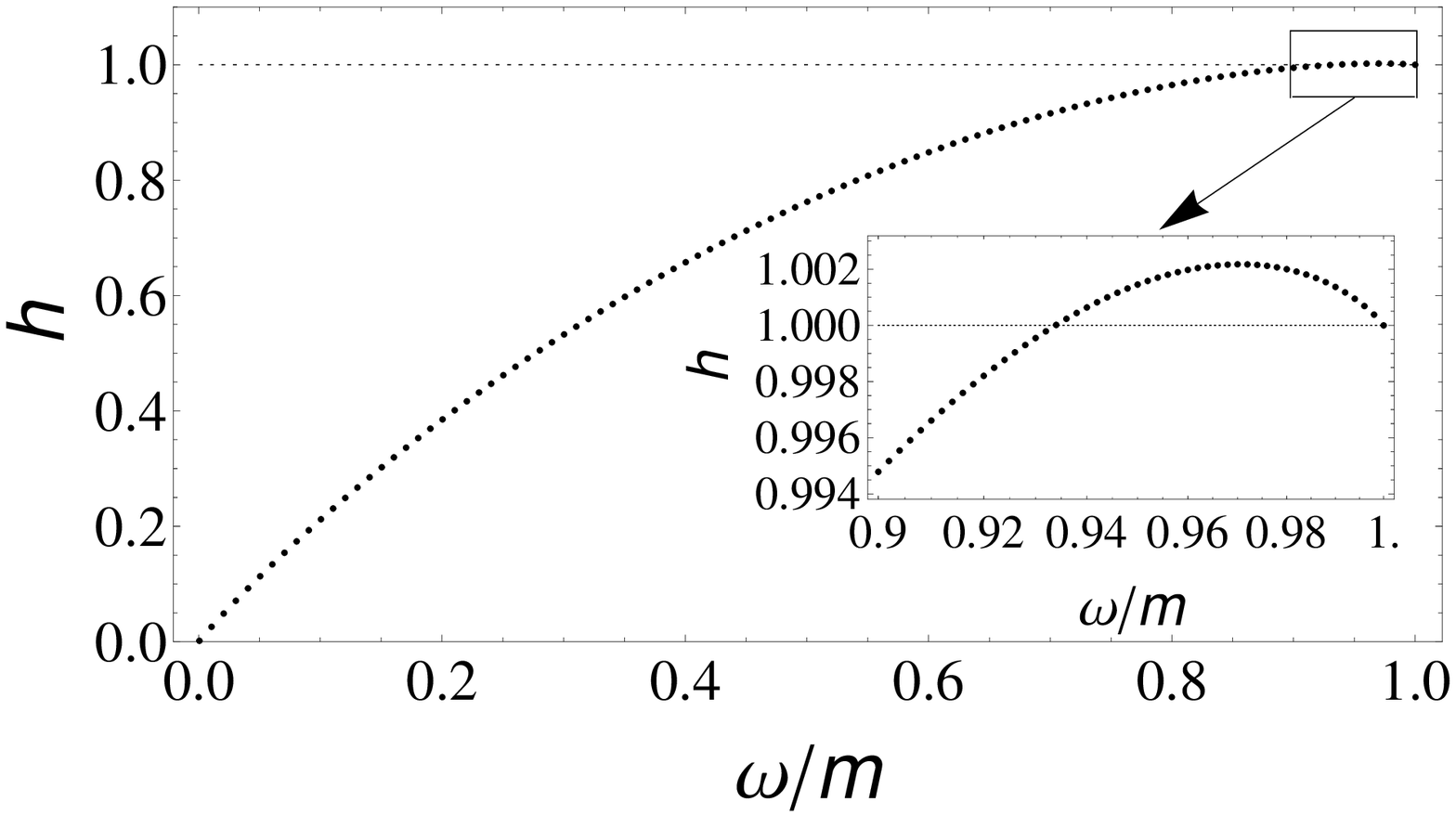}
\includegraphics[clip,width=0.47\textwidth]{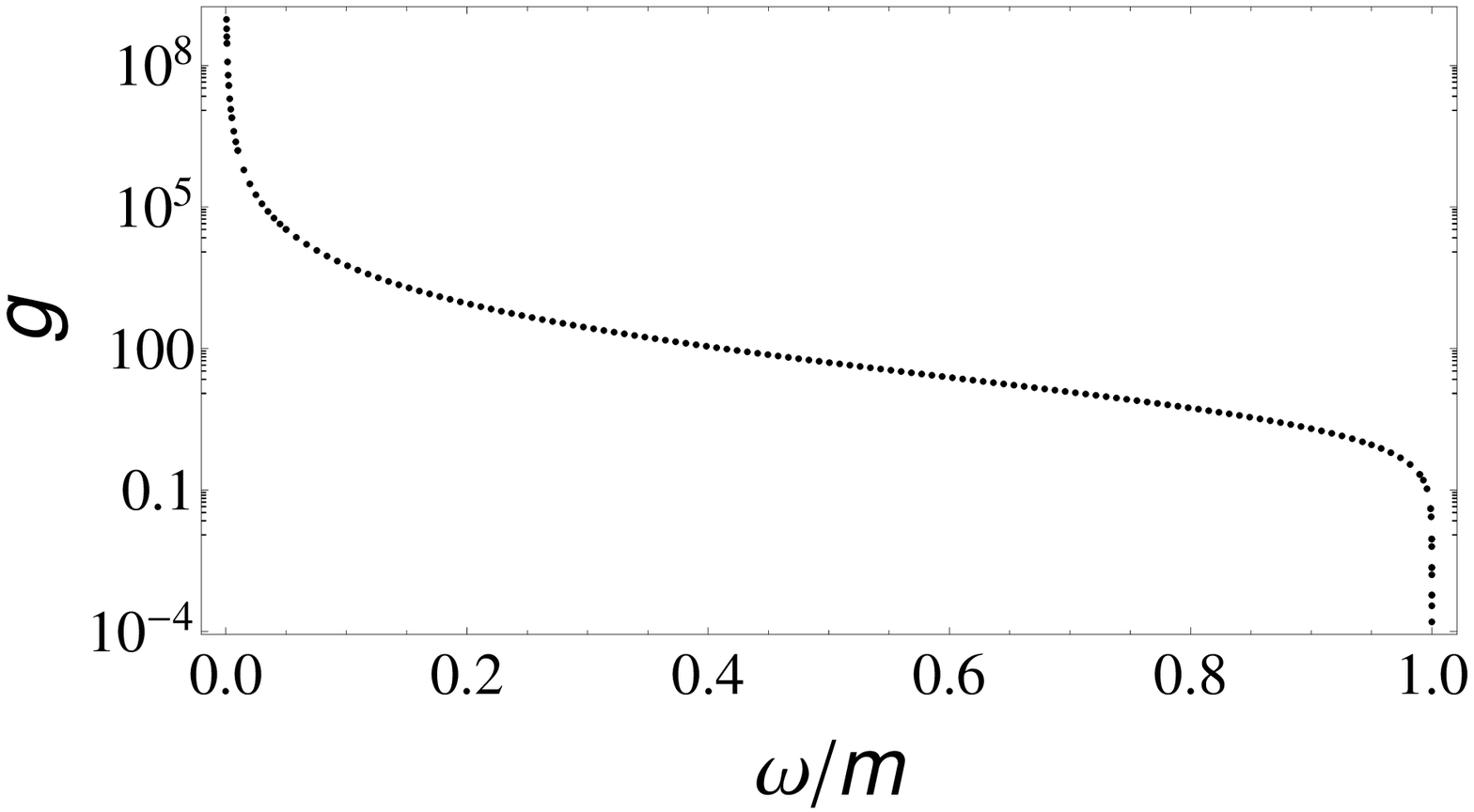}
\caption{\label{Fig:ghF}{\em Upper panel.} The ratio $h = \rho/m
\sigma $ as a function of $\omega$. When $h > 1$, SUSY Q-walls are
absolutely unstable and decay into particles of mass $m$.
{\em Lower panel.} The function $g = \sigma/\sigma_{\rm cr}$ as a function of
$\omega$.}
\end{center}
\end{figure}

Besides we introduce, in analogy with Eq.~\eqref{sigmax}, the
function
\begin{equation}
g(\omega) =\sigma/ \sigma_{\rm cr}~. \label{eq:sigmaxF}
\end{equation}
We plot $g(\omega)$ in the lower panel of Fig.~\ref{Fig:ghF}.
It is interesting to compare Fig.~\ref{Fig:ghF} with Fig.~\ref{Fig:1},
the latter being the result of the calculation within the flat-potential
approximation: The derivative of the total charge with
respect to $\omega$ is negative in the whole range of the allowed
values of $\omega$ for the full potential case. The quantum stability
condition~\eqref{eq:QSta} implies that Q-wall solution is
classically stable in the whole range of $\omega$ (and stable
against fission as well). On the other hand, in the case of the
flat-potential approximation, we have already noted that there exists a
critical value of $\omega$ above the which the derivative changes
sign, signaling a regime of classical instability.


In Fig.~6, we present $\omega$ and $\rho$ as a function of the
charge (points in the plots). Inspired by
Eqs.~\eqref{omegaA}-\eqref{rhoA}, we write $\omega$ and $\rho$ in
the following way:
\begin{eqnarray}
\label{omegaw} && \frac{\omega}{m} = \xi_\omega(\sigma) \left( \frac{\sigma}{\sigma_{\rm cr}} \right)^{\!-1/2} \!, \\
\label{rhow}   && \frac{\rho}{m\sigma_{\rm cr}} = \xi_\rho(\sigma)
\left( \frac{\sigma}{\sigma_{\rm cr}} \right)^{\!1/2} \!.
\end{eqnarray}
We checked that the functions $\xi$ are slowly varying functions
of the charge $\sigma$, and indeed are well approximated by power
functions of the logarithm of the charge:
\begin{equation}
\label{xiw} \xi(\sigma) = [a + b \log_{10} (\sigma/\sigma_{\rm
cr})]^q.
\end{equation}
They parameterize the deviation from the simple power-laws
obtained in the flat-potential approximation. In Table I, we
report the values of the coefficients $a$, $b$, $q$ found by
least-squaring the numerical data for large charges. We also show
the maximum percentage error of the functions $\xi$ with respect
to their numerical values. The dotted lines in Fig.~6 are indeed
the approximate expressions~\eqref{omegaw}-\eqref{rhow}.


\begin{table}[t]
\caption{Nonlinear fit of the functions $\xi$ defined in
Eqs.~(\ref{omegaw})-(\ref{rhow}), (\ref{xiw}), (\ref{deltaw}) and
(\ref{phi0w}). The quantity $\mbox{Err\%}$ represents the maximum
percentage error of the functions $\xi$ with respect to their
numerical values in the range $\sigma \in [10^4 \sigma_{\rm cr},
9.5 \times 10^{8} \sigma_{\rm cr}]$.}

\vspace{0.5cm}

\begin{tabular}{lllllllll}

\hline \hline

&$$        &$~~~~~~~~a$  &$~~~~b$    &$~~~~~\,q$   &$\;\:\mbox{Err\%}$ \\

\hline

&$\omega$  &$~~-0.024$   &$~2.033$   &$~~~~~\,1$   &$~0.06 \%$         \\
&$\rho$    &$~~-5.109$   &$~5.288$   &$~~~1.031$   &$~0.08 \%$         \\
&$\delta$  &$~~-1.542$   &$~1.502$   &$\:-0.978$   &$~0.03 \%$         \\
&$\phi_0$  &$~~-13.961$  &$~4.486$   &$~~~0.027$   &$~0.06 \%$         \\

\hline \hline

\end{tabular}
\end{table}


\begin{figure}[t]
\begin{center}
\includegraphics[clip,width=0.47\textwidth]{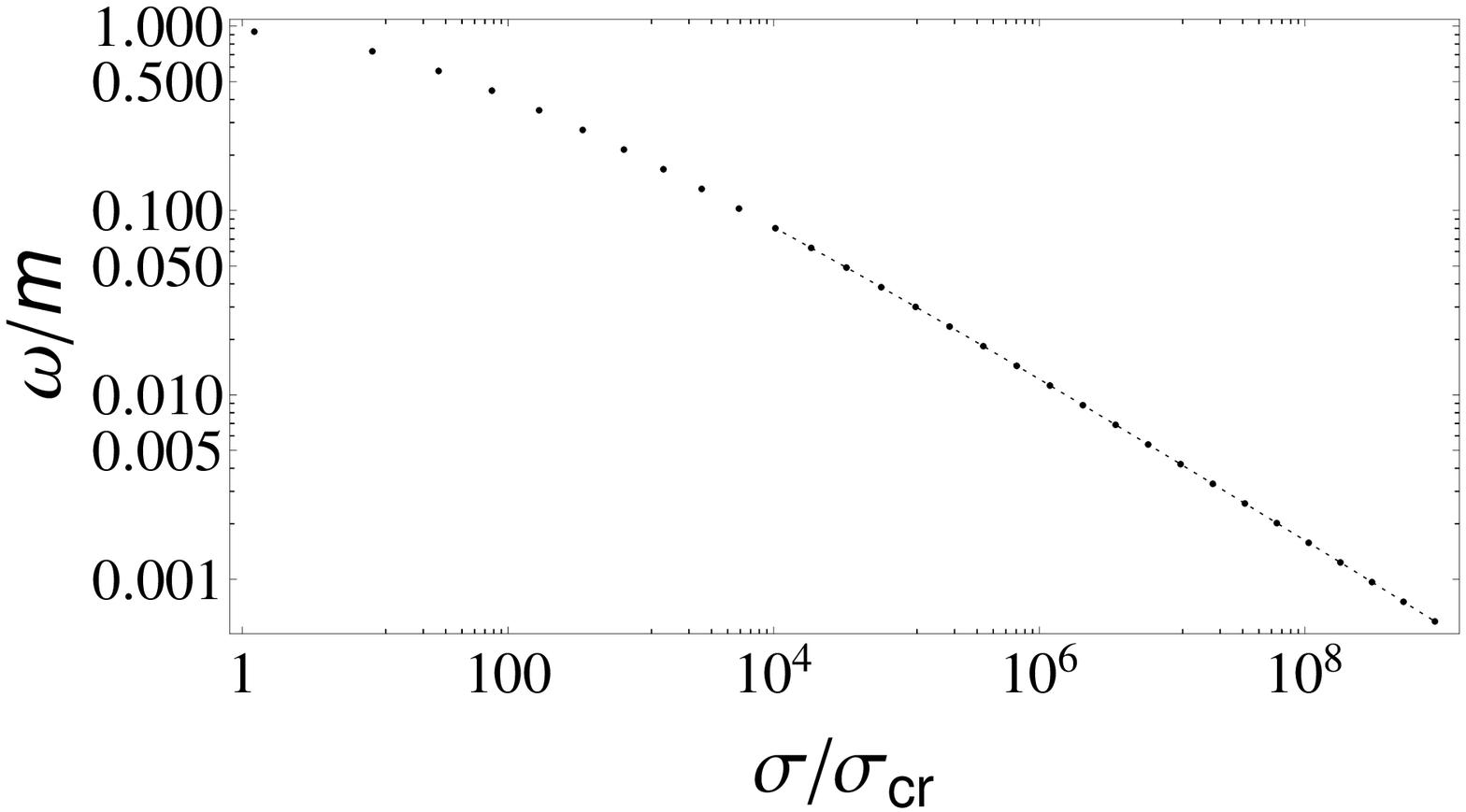}
\includegraphics[clip,width=0.47\textwidth]{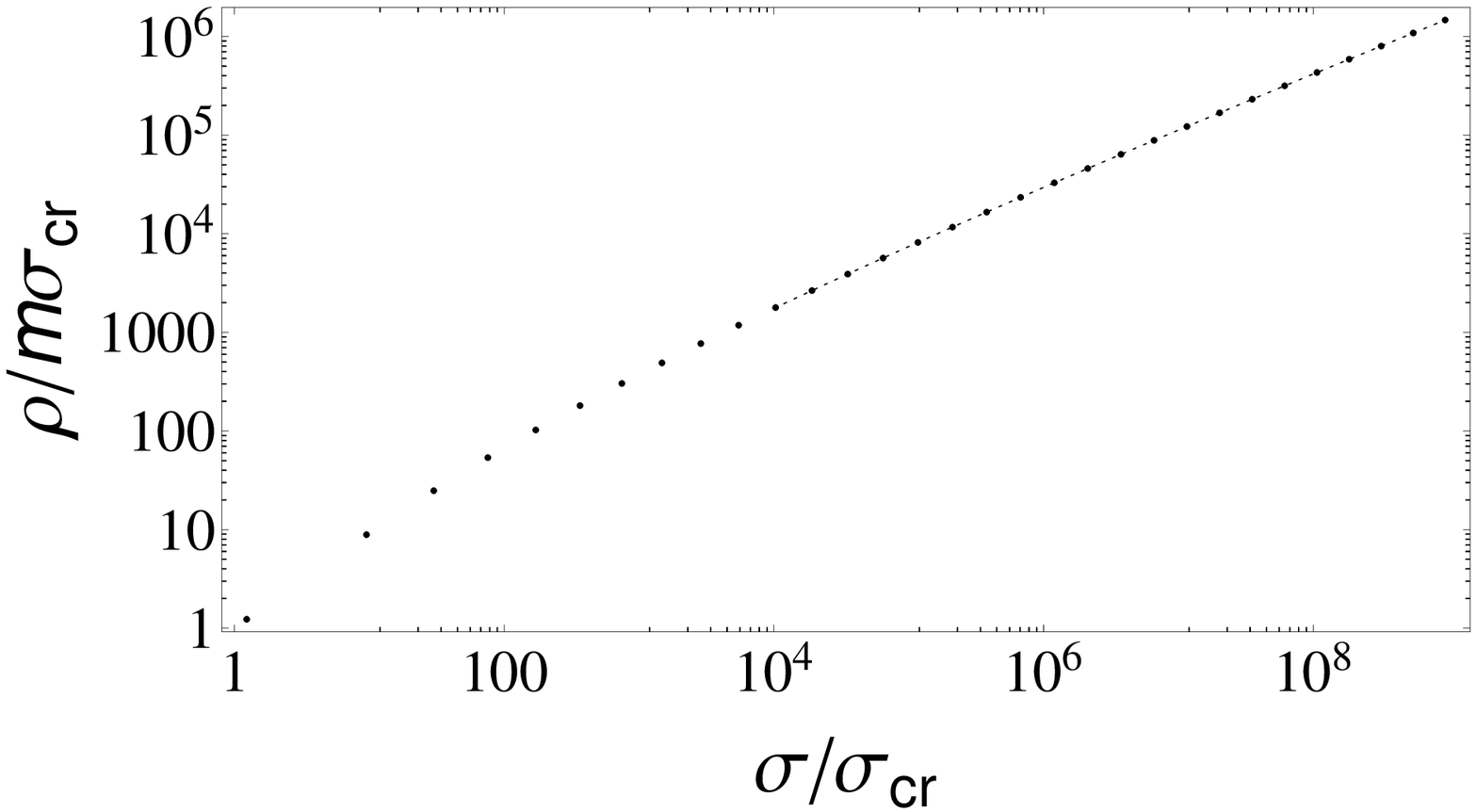}
\caption{The parameter $\omega$ (upper panel) and the surface
energy density $\rho$ (lower panel) as a function of the surface
charge $\sigma$. Dotted lines represent the approximate
expressions of $\omega$ and $\rho$ for large values of the
charge.}
\end{center}
\end{figure}


\begin{figure}[t]
\begin{center}
\includegraphics[clip,width=0.47\textwidth]{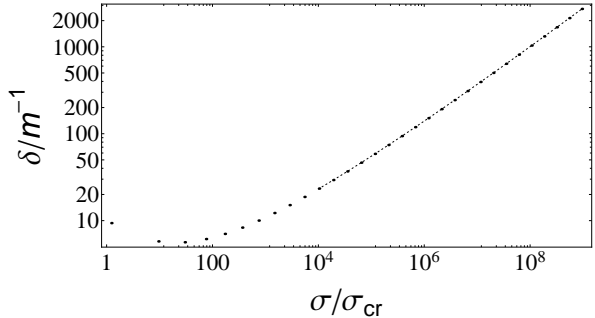}
\caption{The total width of a SUSY Q-wall $\delta$ (points), as a
function of the charge, together with its approximate expression
for large charges (dotted line).}
\end{center}
\end{figure}


\begin{figure}[t]
\begin{center}
\includegraphics[clip,width=0.47\textwidth]{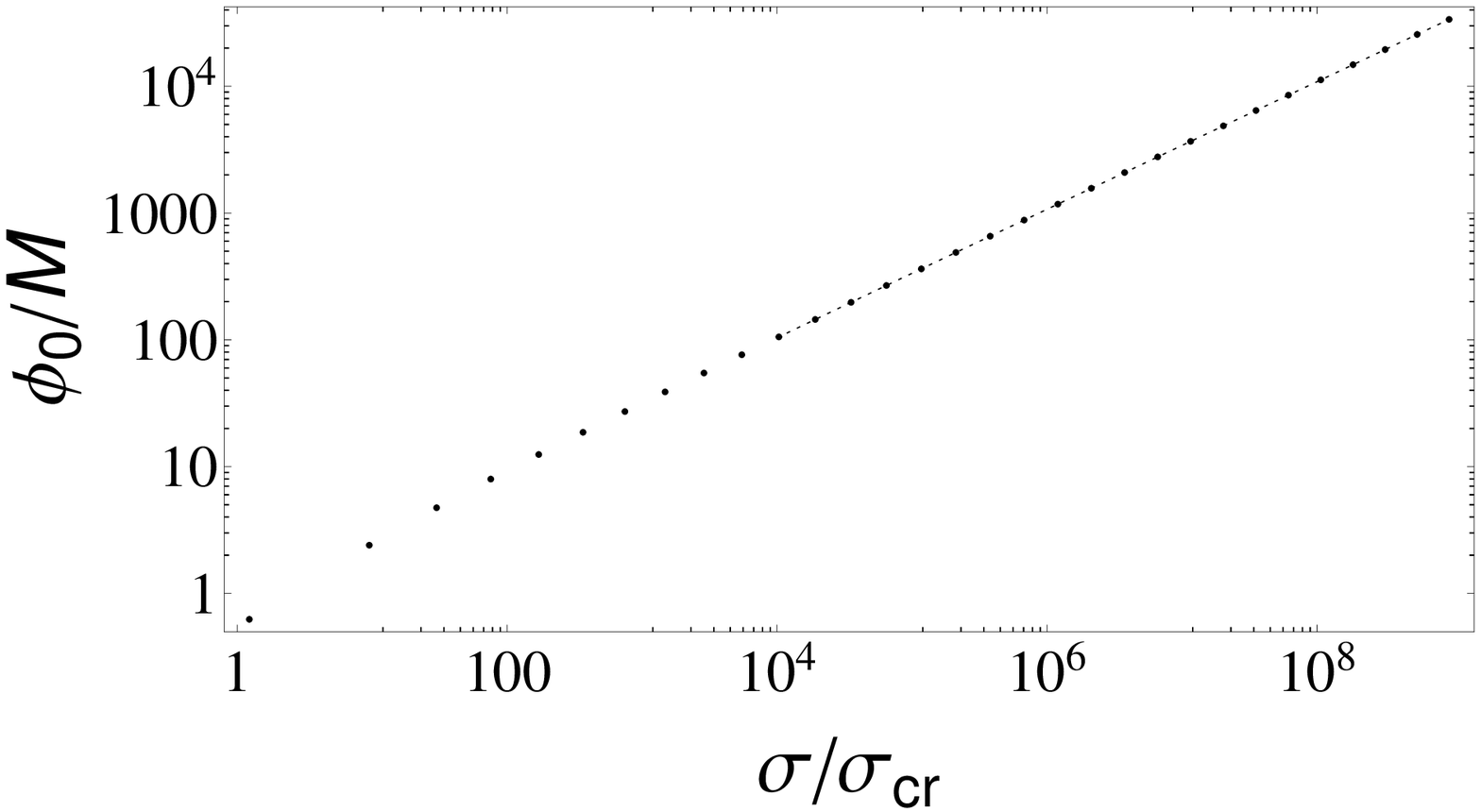}
\includegraphics[clip,width=0.47\textwidth]{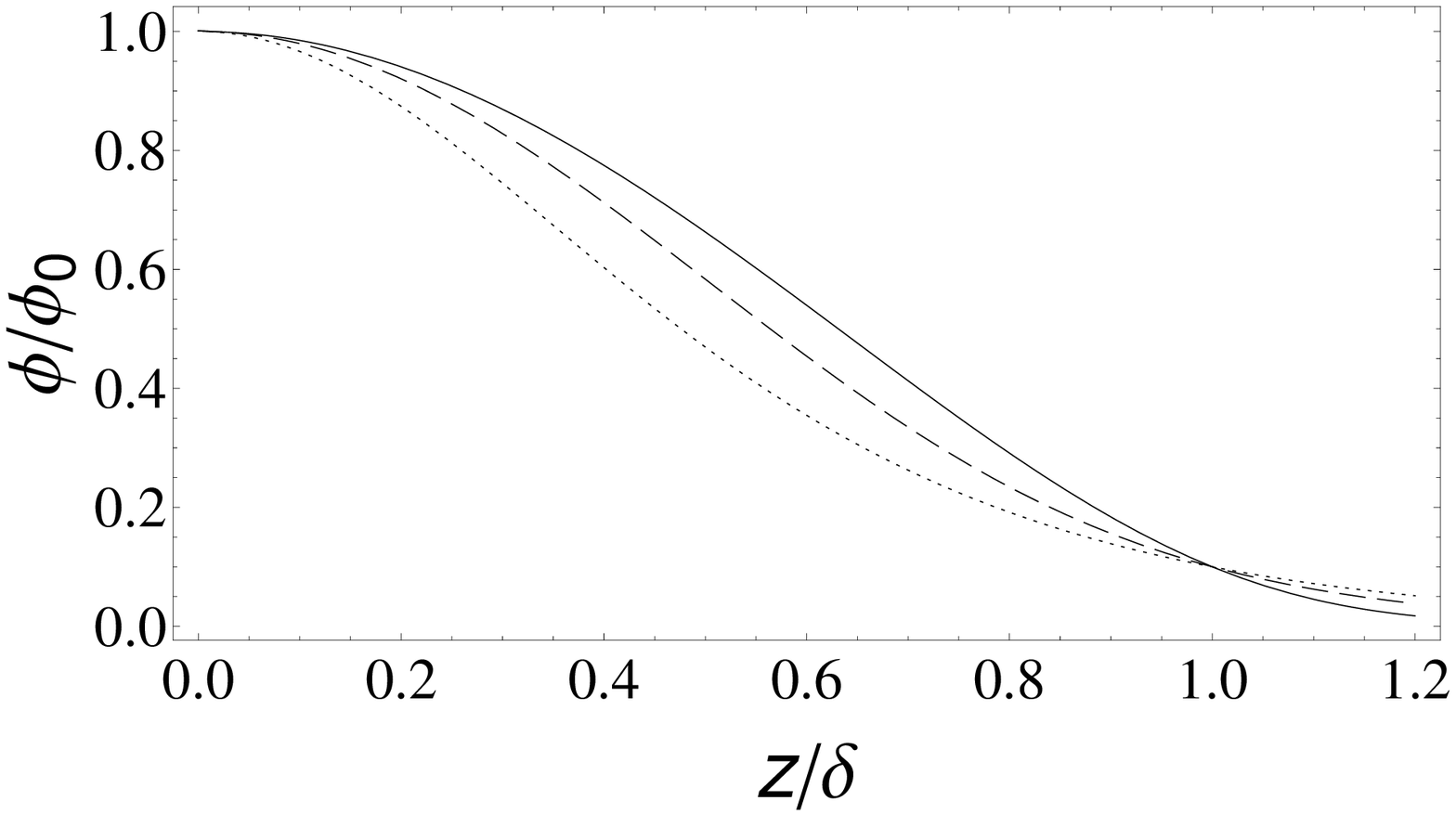}
\caption{{\it Upper panel.} $\phi_0$ as a function of the charge
(points) with its approximate expression for large charges (dotted
line). {\it Lower panel.} SUSY Q-wall profile for different values
of the charge: $\sigma = 10^3 \sigma_{\rm cr}$ (continuous line),
$\sigma = 50 \sigma_{\rm cr}$ (dashed line), $\sigma = \sigma_{\rm
min} \simeq 1.23 \sigma_{\rm cr}$ (dotted line).}
\end{center}
\end{figure}


For the full potential case, we are no longer able to derive
analytical Q-wall profiles. Therefore,
we must use some phenomenological definition for the Q-wall width,
$\delta$, in analogy with our previous definition~\eqref{deltaDef}.
It is convenient to adopt the following definition:
\begin{equation}
\label{deltaFulldef} \phi(\delta) \equiv 0.1 \, \phi_0~.
\end{equation}
In Fig.~7, we show $\delta$ as a function of the charge (points in
the plot) with its approximate expression for large charges
(dotted line):
\begin{equation}
\label{deltaw} \frac{\delta}{m^{-1}} = \xi_\delta(\sigma) \left(
\frac{\sigma}{\sigma_{\rm cr}} \right)^{\!1/2} \!.
\end{equation}
In Fig.~8, we finally show $\phi_0$ as a function of the charge
(points in the plot) with its approximate expression for large
charges (dotted line):
\begin{equation}
\label{phi0w} \frac{\phi_0}{M} = \xi_\phi(\sigma) \left(
\frac{\sigma}{\sigma_{\rm cr}} \right)^{\!1/2} \!.
\end{equation}
Also shown in the figure is the Q-wall profile for different values
of the charge.


\section{Comparison of supersymmetric Q-walls and Q-balls}

Up to now, we have considered two dimensional Q-wall
configurations whose surface area is then indefinitely large. In
real situations, however, the surface area cannot be infinite. To
retain all the above results about Q-walls, we must impose that
the typical longitudinal dimension of a wall,
\begin{equation}
\label{L} L \equiv \sqrt{A},
\end{equation}
is much greater than the typical transverse dimension $\delta$
defined in Eq.~\eqref{deltaFulldef}.

\subsection{Energetics of Finite-size Q-walls}

In the case of {\it finite-size Q-walls}, we can
associate to a Q-wall a finite charge $Q$ and a finite energy
$E_w$ in the following way:
\begin{equation}
\label{QEw} Q = \sigma L^2, \;\;\;\, E_w = \rho L^2,
\end{equation}
whenever the {\em feasibility condition} is satisfied:
\begin{equation}
\label{alpha} \alpha \equiv \frac{L}{\delta} \gg 1.
\end{equation}
(In the following we assume, for definiteness, that $\alpha \geq
10$). The ideal (but unphysical) case of infinite wall surface is
recovered for $L \rightarrow \infty$ or, equivalently, for $\alpha
\rightarrow \infty$.

Before proceeding further, let us observe that for a given $L$
there exists a minimum charge defined by
\begin{equation}
\label{Qminwall} Q^{\rm min}_{w} \equiv \sigma_{\rm min} L^2.
\end{equation}
This immediately follows from Eq.~\eqref{sigmamin}. It is clear
that for fixed values of $\alpha$, the various quantities defining
a Q-walls, e.g. energy, total width, etc., will depend now on both
$\alpha$ and $Q$:
\begin{equation}
E_w = E_w(Q,\alpha), \;\;\; \delta = \delta(Q,\alpha), \;\;\;
\mbox{etc.},
\end{equation}
while the minimum charge only on $\alpha$:
\begin{equation}
Q^{\rm min}_w = Q^{\rm min}_{w}(\alpha).
\end{equation}
(In the case of infinite Q-walls the above quantities depend only
on the surface charge $\sigma$, while $\sigma_{\rm min}$ is
fixed.)

As an example, let us write the expression of the energy and
charge of a {\it finite-size SUSY Q-wall} in the flat-potential
approximation:
\begin{eqnarray}
\label{EwA} && \frac{E_w}{m Q_{\rm cr}} \simeq \pi^{3/4} \left( \frac{Q}{Q_{\rm cr}} \right)^{\!3/4} \alpha^{1/2}, \\
\label{QwA} && \frac{Q}{Q_{\rm cr}} \simeq \pi \left( \frac{\sigma}{\sigma_{\rm cr}} \right)^{\!2} \alpha^2,
\end{eqnarray}
where we remind that $Q_{\rm cr} \equiv \Lambda/m^4$. The above
equations are valid in the limit $\sigma \gg \sigma_{\rm cr}$. In
particular, for Eq.~(\ref{EwA}) this means $Q \gg (\sigma_{\rm
cr}/\sigma_{\rm min}) \, Q^{\rm min}_w \simeq Q^{\rm min}_w/3$,
where the minimum charge is
\begin{equation}
\label{QwminA} \frac{Q^{\rm min}_w}{Q_{\rm cr}} \simeq \pi \left(
\frac{\sigma_{\rm min}}{\sigma_{\rm cr}} \right)^{\!2} \alpha^2
\simeq 9 \pi \alpha^2.
\end{equation}
Comparing Eq.~(\ref{EwA}) with Eq.~(\ref{Eball1}), namely the
energy of a SUSY Q-wall with the energy of a SUSY Q-ball
possessing the same charge Q, we find that
\begin{equation}
\label{r} \frac{E_w}{E_b} \simeq \frac{3}{4\sqrt{2}\pi^{1/4}} \,
\alpha^{1/2} \simeq 0.40 \, \alpha^{1/2},
\end{equation}
valid for $Q \gg \mbox{max}[Q^{\rm min}_b, Q^{\rm min}_w]$, where
$Q^{\rm min}_b$ is the minimum charge required for stable Q-balls.
Since we are assuming $\alpha \geq 10$, we find that $E_w > E_b$
for $\alpha \in [10,\infty]$.

We want now to analyze the more realistic case of full
supersymmetric potential giving rise to Q-walls and Q-balls. To
this end, in Fig.~9, we show the ratio between the energy of a
SUSY Q-wall $E_w$ and the energy of a SUSY Q-ball $E_b$, as a
function of the charge $Q$ and for different values of $\alpha$.
We find that $Q^{\rm min}_w > Q^{\rm min}_b \simeq 504 Q_{\rm cr}$
for all $\alpha \geq 10$, and that large planar Q-walls ($\alpha
\gtrsim 30$) are more energetic than Q-balls with the same charge.


\begin{figure}[t]
\begin{center}
\includegraphics[clip,width=0.47\textwidth]{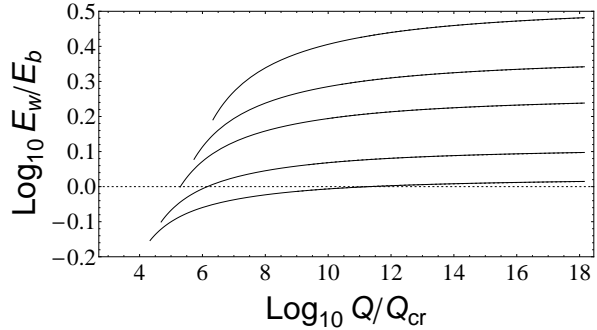}
\caption{The ratio between the energy of a SUSY Q-wall $E_w =
\sigma L^2$ and the energy of a Susy Q-ball $E_b$, as a function
of the charge $Q$ and for different values of $\alpha = L/\delta$.
From down to top: $\alpha = 10,15,30,50,100$. The horizontal
dotted line is $E_w = E_b$. In the case of walls, the charge is
$Q=\sigma L^2$, where $\sigma$ is the surface energy density and
$L$ the longitudinal dimension of the wall. In a Q-wall
configuration, $L$ is always much greater than the transverse
dimension $\delta$.}
\end{center}
\end{figure}


However, it is very interesting to observe that there exist Q-wall
configurations with moderate values of $\alpha$ ($10 \leq \alpha
\lesssim 30$) that are {\it less} energetic of the corresponding
Q-balls if the charge is in the range $Q^{\rm min}_w \leq Q \leq
Q^*$, where $Q^{\rm min}_w$ and $Q^*$ as a function of $\alpha$
are shown in Fig.~10, $Q^*$ being defined by the equality
\begin{equation}
\label{Qstar} E_w(Q^*) \equiv E_b(Q^*).
\end{equation}
Stated in other words, we find that for charges in the range
\begin{equation}
2.16 \times 10^4 Q_{\rm cr} \leq Q \leq 2.06 \times 10^{11} Q_{\rm cr}
\end{equation}
(the lower and upper limits correspond to $Q = Q^{\rm min}_w$ and
$Q = Q^*$ evaluated at $\alpha=10$) there always exists a Q-wall
with $10 \leq \alpha \lesssim 30$ such that
\begin{equation}
E_w < E_b.\label{eq:70}
\end{equation}

\subsection{Estimate of the Lifetime of a Finite-size Q-wall}

Equation~\eqref{eq:70} establishes that, for particular values of
the charge, finite-size Q-walls are energetically favored over
Q-balls with the same value of the charge. The straightforward
interpretation of this result would be that finite-size,
non-topological Q-walls can possibly be the ground state of the
scalar theory defined by Lagrangian~(\ref{eq:Lagr1}). However, due
to their finiteness, such Q-walls are {\em dynamically} unstable.
It is clear, indeed, that under the action of their own tension,
finite-size Q-walls with charge $Q$ will eventually shrink. The
typical size of the Q-wall, when this happens, is found by
equating its energy to that a Q-ball with same charge $Q$:
\begin{equation}
\label{EwallL} E_w(L_{\rm decay},Q) = E_b(Q).
\end{equation}
Using the flat-potential approximation, it is straightforward to
obtain
\begin{equation}
\label{Ldecay} L_{\rm decay} =\frac{2\sqrt{2\pi}}{3} \left(
\frac{Q}{\Lambda} \right)^{1/4} \, ,
\end{equation}
where we used Eqs.~(\ref{rhoA}) and (\ref{QEw}) and
Eq.~(\ref{Eball1}).
%
%

In order to estimate the lifetime of a finite-size Q-wall, we
follow the approach of Ref.~\cite{MacKenzie:2001av}. We introduce
an effective Lagrangian defining the dynamics of a Q-wall of size
$L$:
\begin{equation}
\label{LagrangianWall} \mathcal{L}_w = \frac12 \, m_w(L) \,
\dot{L}^2 - V_w(L),
\end{equation}
where $m_w(L) = E_w(L)$ and $V_w(L) =E_w(L)$ are the effective
mass and tension of the wall, respectively.
%
%
The equation of motion for $L(t)$ is then
\begin{equation}
\label{L(t)} \frac12\, L \dot{L}^2 + L  = L_i \, ,
\end{equation}
where we have indicate with $L_i$ the wall scale-length when the wall
begins to shrink at the time $t_i$. The lifetime of the wall, i.e.
the time interval between $t_i$ and the time when the wall decays
into a Q-ball, $t_{\rm decay} \equiv t(L_{\rm decay})$, is given
by integrating Eq.~(\ref{L(t)}):
\begin{equation}
\label{tauWall} \tau_w = t_{\rm decay} \, - \, t_i =
\frac{\pi}{2\sqrt{2}} \, L_i \, f(L_{\rm decay}/L_i),
\end{equation}
where
\begin{equation}
\label{f} f(x) \equiv \frac{2}{\pi} \left[ \sqrt{x( 1 - x)} +
\arccos \sqrt{x} \, \right] \! .
\end{equation}
We note that $f(x)$ is a decreasing function of $x$ such that
$f(0) = 1$ and $f(1) = 0$. Consequently, we have
\begin{equation}
\label{tauWall2} \tau_w \leq \frac{\pi}{2\sqrt{2}} \, L_i \simeq
1.11 \, L_i \, .
\end{equation}

\subsection{Phase diagram}


\begin{figure}[t]
\begin{center}
\includegraphics[clip,width=0.47\textwidth]{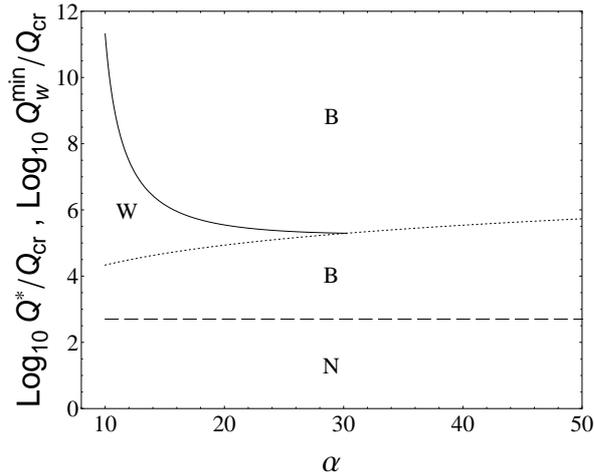}
\caption{\label{Fig:PD} Phase diagram of non-topological
solitons in a model with broken supersymmetry via low-energy gauge
mediation. The regions denoted by B (W) and N correspond, respectively,
to domains where Q-balls (finite-size Q-walls) and unbound quanta are stable (meta-stable).
The dotted line represents the minimum charge possessed by a finite-size, SUSY
Q-wall, $Q^{\rm min}_w$, at the varying of the ratio $\alpha$ between longitudinal
and transverse dimension of the wall.
The continuous line is the charge $Q^*$
defined by $E_w(Q^*)=E_b(Q^*)$, as a function of
$\alpha$.
The dashed line is the the minimum charge
required for a stable SUSY Q-ball, $Q^{\rm min}_b \simeq 504
Q_{\rm cr}$.
W is a meta-stability region for Q-walls: Although finite-size Q-walls are less energetic than Q-balls in this
region, finite-size effects cause Q-walls to decay into
Q-balls with the same value of the charge.}
\end{center}
\end{figure}


The previous discussions can be summarized in Fig.~\ref{Fig:PD},
where the regions denoted by B, W, and N correspond, respectively,
to domains in which Q-balls, Q-walls, and unbound quanta are
stable (N corresponds to {\em normal phase}, as it is customary to
define the unpaired phase for superfluids and superconductors). In
particular, the region denoted by W is bounded by two transition
lines from a ground state made of Q-walls to the ground state made
by Q-balls. In this region, Eq.~\eqref{eq:70} is satisfied.
However, since finite-size Q-walls are dynamically unstable, it is
more appropriate to call the region W as the {\em meta-stability
region} of Q-walls.

This region has to be understood as follows: If a point $(\alpha, Q)$
lies in the region W, then the energy of the corresponding planar-symmetric, finite-size soliton
is lower than the energy of a spherical-symmetric soliton possessing the same charge.
However, that planar-symmetric soliton has a finite lifetime (due to its dynamical instability)
and its charge-conserving decay is a spherical-symmetric soliton.


\section{Conclusions}

We have investigated a particular class of
non-topological solitons possessing planar symmetry, known as
Q-walls, in the context of a supersymmetric particle physics model
with supersymmetry broken by low-energy gauge mediation.

On general grounds, solutions with spherical symmetry (Q-balls)
are expected to be favored over the ones with planar symmetry.
Q-walls, then, might be interpreted as excited states of Q-balls.
The study of such excited states is interesting for they may form
as intermediate states when dynamics is involved --e.g. in
processes of Q-balls collision, as well as in soliton production
via fragmentation of the Affleck-Dine condensate.

Firstly, we have discussed general analytic properties of
Q-wall configurations arising in field theories
invariant under a global $U(1)$ transformation
and without relying on any specific form of the potential
giving rise to them. This analysis
is therefore relevant for any model whose potential allows for
Q-walls.

Secondly, we have analyzed the peculiar properties of Q-walls within a
specific supersymmetryc model, namely that in which supersymmetry is
broken at the quantum level by virtue of low-energy gauge mediation.
The use of an approximate form of the two-loop potential that breaks supersymmetry
has led us to simple analytic results, in agreement with previous works on
this subject.

Also, we have studied Q-wall solutions considering the exact form
of the supersymmetric potential. Only numerical results have been
obtained which are, however, in fairly close agreement with
those derived in the approximate case.

Finally, we have compared the energy of a finite-size Q-wall with
that of a Q-ball, at fixed charge $Q$. The result is summarized in
the ``phase diagram'' of Fig.~\ref{Fig:PD} in which a region
exists --region W in Fig.~10-- where Q-walls are less energetic
than Q-balls.

Such a region is, however, a meta-stability region since an
effective analysis of stability has revealed that finite-size
Q-walls are unstable with respect to the decay into Q-balls.

\acknowledgments The work of M.~R. is supported by JSPS under the
contract number P09028.


\end{document}